\documentclass[onecolumn,prd,showpacs,nofootinbib]{revtex4}
\usepackage{graphicx}
\usepackage{textcomp}
\usepackage[usenames,dvipsnames]{color}
\usepackage{amsmath}
\usepackage{amssymb}
\usepackage{slashed}
\usepackage{amsmath}
\usepackage{latexsym}
\usepackage[all]{xy}
\usepackage{color}
\usepackage{hyperref}
\include{amslatex}

\numberwithin{equation}{section}

\begin{document}

\title{Phenomenology of effective geometries from quantum gravity}

\author{Ricardo Gallego Torrom\'e}
\affiliation{Departamento de Matem\'atica, Universidade Federal de S\~ao Carlos, Rodovia Washington Lu\'is, km 235, SP, Brazil}

\author{Marco Letizia and Stefano Liberati}
\affiliation{SISSA, Via Bonomea 265, 34136 Trieste and INFN Sezione di Trieste}

\begin{abstract}
In a recent paper \cite{Assaniousssi:2014ota} a general mechanism for emergence of cosmological space-time geometry from a quantum gravity setting was devised and departure from standard dispersion relations for elementary particle were predicted. We elaborate here on this approach extending the results obtained in that paper and showing that generically such a framework will not lead to higher order modified dispersion relations in the matter sector. Furthermore, we shall discuss possible phenomenological constraints to this scenarios showing that space-time will have to be by nowadays classical to a very high degree in order to be consistent with current observations.
\end{abstract}

\maketitle

\section{Introduction}

The quest for a quantum gravity theory has been at the forefront of research in theoretical physics for more than fifty years by now. This quest has lead to the development of admirable theories based on sophisticated mathematics and physical intuition. We have nowadays a broad landscape of theories aiming at describing gravity at the Planck scale, from theories of everything like string theory to pure quantum gravity approaches such as Loop Quantum gravity (in its different declinations such as Canonical, Covariant and group field theory quantisations) \cite{Ashtekar:2008zu, Ashtekar:2008ay} and Causal Dynamical Triangulation \cite{Ambjorn:2013tki,Ambjorn:2013hma} or Causal Set Theory \cite{Bombelli:1987aa,Surya:2011yh} just to name a few.

It is however undoubtful that in developing these theories researchers have been dealing with a challenge that physics did not have to face in its early centuries of development; the lack of empirical guidance. While the difficulty to probe Planck scale physics with current observational and experimental capacities has induced some practitioners to the conjecture that we are entering a ``post-empirical" phase of theoretical physics, there are reasons to believe that not such dramatic departure from the standard ``scientific method" are needed. The past decades have in fact lead to a new field broadly called ``quantum gravity phenomenology" which has shown as in several situations some information can indeed be extracted from sub-Planckian observations about the microscopic structure of space-time.  While it would be inappropriate to dwell on these phenomenology in the present manuscript we just direct the reader to several recent reviews of the issue \cite{Liberati:2013xla,AmelinoCamelia:2008qg,Hossenfelder:2010zj}.

In all these attempts to test QG scenarios a crucial issue is the continuous and classical limit of these theories. Noticeably these two limits do not need to coincide and their order does matter in extracting phenomenological consequences. A very illustrative and general example in this sense was presented in a recent paper \cite{Assaniousssi:2014ota}, where a general mechanism for emergence of cosmological space-time geometry from a quantum gravity setting was discussed. Such a mechanism was based on very general assumption about the existence of quantum gravitational degrees of freedom to be described in terms of a state $\Psi_0$ in a Hilbert space ${\cal H}_G$, which can be considered ``heavy" compared to the matter degrees of freedom (in the Born-Oppenheimer sense). Assuming negligible back reaction of the matter degrees of freedom on the gravitational ones, one can suitably trace away the latter so to obtain an effective continuous space-time characterised by a dimensionless parameter measuring its degree of classicality and in principle energy dependent (from here the name of rainbow geometry \cite{Lafrance:1994in,Magueijo:2002xx}). In \cite{Assaniousssi:2014ota} a low momentum approximation was used to derive a modified dispersion relation showing a different limit speed of propagation for elementary particles plus higher order terms in momentum.

In this paper we shall revise the analysis of \cite{Assaniousssi:2014ota} and discuss its phenomenological implication. In particular we shall show that when an exact calculation is performed while  a rainbow (energy dependent) geometry indeed emerges from the framework devised in \cite{Assaniousssi:2014ota} the modified dispersion relation there obtained in the low momentum limit is indeed exact, i.e.~no momentum dependent departures from Lorentz invariance appears. Finally, we shall discuss in detail what kind of phenomenological constraints can be derived for such modified dispersion relation and which perspectives for improvements are offered by this general treatment.

\section{Cosmological spacetimes from quantum gravity}\label{cosmst}

Following \cite{Assaniousssi:2014ota} we shall start by considering a massive scalar field $\phi$ minimally coupled to gravity. When performing a separation of the homogeneous and the inhomogeneous degrees of freedom, as in the analysis presented in \cite{Ashtekar:2009mb}, one can describe the classical dynamics for a mode $k$ of the field (up to second order) via an Hamiltonian of the form
\begin{equation}\label{classicalHamiltonianforkmode}
H_{\vec{k}}=\,H_0-\,\frac{1}{2}\,H^{-1}_0\Big[\pi^2_{\vec{k}}+\,(\vec{k}^2\,a^4+m^2a^6 )\phi^2_{\vec{k}}\,\Big],
\end{equation}
with $H_0$ the Hamiltonian of the homogeneous gravitational degrees of freedom $(a,\pi_a)$ and $(\phi_{\vec{k}},\pi_{\vec{k}})$ are the variables in the phase space of the $\vec{k}$-mode of the scalar field \cite{Ashtekar:2009mb}.
After formal quantization of matter and gravity, \eqref{classicalHamiltonianforkmode} can be used to define a Schr\"odinger-like equation \cite{Assaniousssi:2014ota}.
The formal procedure followed in \cite{Ashtekar:2009mb,Assaniousssi:2014ota} in the quantization of the Hamiltonian \eqref{classicalHamiltonianforkmode} is to consider the product space $\mathcal{H}=\,\mathcal{H}_G\otimes \mathcal{H}_{m}$, with $\mathcal{H}_G$ the Hilbert space for the gravitational degrees of freedom and $\mathcal{H}_m$ the Hilbert space for matter, represented here by the scalar field\footnote{This procedure is quite general and does not rely on the specific form of $\mathcal{H}_G$. Nevertheless one should be able to describe the gravitational sector at the quantum level by mean of a Hilbert space as well as perform a factorization between the gravitational and matter Hilbert spaces. These are not trivial requirements and they are among the assumptions on which this model is based.}. Assuming negligible back reaction on the gravitational part by matter one can then trace away the gravitational degrees of freedom described by the state $\psi_0\in \mathcal{H}_G$, and so formally obtain an effective Hamiltonian for the matter sector

\begin{equation}\label{quantumfundamentalHamiltonian}
\hat{H}^{\rm traced}_{\vec{k}}=\,\frac{1}{2}\,\Big[\langle\psi_0|\hat{H}^{-1}_0|\psi_0\rangle\,\hat{\pi}_{\vec{k}}^2+
\,\langle\psi_0|\hat\Omega(\vec{k},m)|\psi_0\rangle\,\hat{\phi}_{\vec{k}}^2\Big],
\end{equation}
with
\begin{equation}\label{omega}
\hat\Omega(\vec{k},m)=\,\vec{k}^2\widehat{H^{-1}_0a^4}+\,m^2\widehat{H^{-1}_0a^6}\,.
\end{equation}

The Hamiltonian \eqref{quantumfundamentalHamiltonian}  is similar to the one of a quantum scalar field in a \textit{classical} FRLW space-time given by the following line element
\begin{equation}\label{part}
\bar{g}_{\mu\nu} dx^\mu dx^\nu=-\bar{N}^2 dt^2+\bar{a}^2\left(dx^2+dy^2+dz^2\right).
\end{equation}
The quantum Hamiltonian of a $\vec{k}$-mode is then given by
\begin{align}
\hat{H}^{\rm eff}_{\vec{k},m}=\,\frac{1}{2}\,\frac{\bar{N}}{\bar{a}^3}\,\Big[ \hat{\pi}^2_{\vec{k}}+\,(\vec{k}^2\bar{a}^4+\,m^2\bar{a}^4)\hat{\phi}^2_{\vec{k}}\Big].
\label{quantumHamiltonianforkmodeFRW}
\end{align}
Following \cite{Assaniousssi:2014ota} one can then use the formal analogy between the traced, QG-derived, Hamiltonian \eqref{quantumfundamentalHamiltonian} and the quantum Hamiltonian \eqref{quantumHamiltonianforkmodeFRW}. The matching of the two operators requires that the relations \cite{Assaniousssi:2014ota}
\begin{equation}
\label{equ_scfac}
\bar{a}^6+\frac{\vec{k}^2}{m^2}\bar{a}^4-\delta=0, \quad\mbox{and} \quad
\frac{\bar{N}}{\bar{a}^3}=\,\langle \psi_0|\hat{H}^{-1}_0|\psi_0\rangle \, ,
\end{equation}
are simultaneously satisfied, the parameter $\delta$ defined as
\begin{equation}
\delta:=\frac{\langle \hat\Omega(\vec{k},m) \rangle }{m^2 \langle \hat{H}^{-1}_0\rangle}.
\end{equation}

As pointed out in \cite{Assaniousssi:2014ota}, in contrast with the standard FLRW metric, in this case the metric \eqref{part} allows for a non-trivial dependence of $\bar{N}$ and $\bar{a}$ on $\vec{k}$, due to the conditions \eqref{equ_scfac}. Thus, modes with different momenta will in general see different space-times properties, hence the name of rainbow geometry.

\section{Exact derivation of the quantum geometry}\label{quantgeomsection}

Let us first of all introduce, for the sake of convenience, the parameters $\eta$ and $\xi$ by
\begin{equation}\label{delta}
\delta=\frac{\vec{k}^2}{m^2}\frac{\langle \widehat{H_0^{-1}a^4}\rangle}{\langle \widehat{H}_0^{-1}\rangle}+\frac{\langle \widehat{H_0^{-1}a^6}\rangle}{\langle \widehat{H}_0^{-1}\rangle}:=\frac{\vec{k}^2}{m^2}\xi+\eta,
\end{equation}
and note that a consequence of the second constraint in \eqref{equ_scfac} is that
\begin{align}
\frac{\partial}{\partial k^j}\,\frac{\bar{N}}{\bar{a}^3}=0,\quad j=1,2,3 \, ,
\label{constraintpartialkNa3}
\end{align}
which derives from the absence of $k$ dependence (no backreaction assumption) of the expectation value of gravitational Hamiltonian.
However, as already anticipated, the functions $\bar{N}$ and $\bar{a}$ will depend on $k$.

The three solutions of the first equation in (\ref{equ_scfac}) can be written in the following form
\begin{equation}\label{solutions}
\begin{split}
\bar{a}^2_n=&-\frac{1}{3}\left[\frac{\vec{k}^2}{m^2}+u_n \left(\frac{\vec{k}^6}{m^6}-\frac{27}{2}\left(\frac{\vec{k}^2}{m^2}\xi+\eta\right)+\frac{1}{2}\sqrt{-27 \Delta}\right)^{1/3}\right.+\\&+\left.\frac{\vec{k}^4/m^4}{u_n \left(\dfrac{\vec{k}^6}{m^6}-\dfrac{27}{2}\left(\dfrac{\vec{k}^2}{m^2}\xi+\eta\right)+\dfrac{1}{2}\sqrt{-27 \Delta}\right)^{1/3}}\right],
\end{split}
\end{equation}
with $n=\{1,2,3\}$ and $u_n=\{1,(1/2)(-1+\imath \sqrt{3}),(1/2)(-1-\imath \sqrt{3})\}$.\\ It is possible to distinguish all the cases studying the discriminant given by
\begin{equation}\label{discr}
\Delta=\left(\frac{\vec{k}^2}{m^2}\xi+\eta\right)\left[4\frac{\vec{k}^6}{m^6}-27\left(\frac{\vec{k}^2}{m^2}\xi+\eta\right)\right].
\end{equation}
It is easy to see that the study of the sign of (\ref{discr}) boils down to the study of the following equation
\begin{equation}\label{discr2}
\frac{4}{27}\frac{\vec{k}^6}{m^6}-\left(\frac{\vec{k}^2}{m^2}\xi+\eta\right)=0.
\end{equation}
This equation is exactly the equation $\dfrac{4}{27}\dfrac{\vec{k}^6}{m^6}=\delta$ that differentiates the two branches in \cite{Assaniousssi:2014ota}.

It can be shown that for $\Delta=0$ (one multiple root and all the roots are real) the solutions $\bar{a}_2^2$ and $\bar{a}_3^2$ collapse to a double root that is real but negative. For $\Delta>0$ (three real distinct roots) all the solutions are real but only $\bar{a}_1^2$ is positive. On the other hand, for $\Delta<0$ (one real root and two complex conjugate roots), $\bar{a}_1^2$ is real and positive while $\bar{a}_2^2$ and $\bar{a}_3^2$ are two complex conjugate solutions.
Hence it turns out that only the first solution $\bar{a}^2_1$ is positive over the possible values of $k,\xi$ and $\eta$. From now on, the real, positive solution $\bar{a}^2_1$ will be denoted by $\bar{a}^2$.

\subsection*{Classical limit}

It is natural to require for classical space-times, that is, space-times where quantum effects are disregarded, to be independent of the momentum label $k$.
This condition is given by solving the constraint
\begin{equation}
\frac{\partial}{\partial \vec{k}} \bar{a}^2(\vec{k},m,\xi,\eta)=0.
\end{equation}
The only real solution of the above equation is $\xi=\eta^{2/3}$, that is the classicality condition given in~\cite{Assaniousssi:2014ota}, and indeed the classical scale-factor is momentum-independent in this limit being given by $\bar{a}^2_0:=\bar{a}^2(\vec{k},m,\eta^{2/3},\eta)=\sqrt{\xi}=\eta^{1/3}$.

\subsection*{Parameter of non-classicality}\label{classicalitysection}

While it is clear that the are two independent parameters associated to the non-classicality of the fundamental Hamiltonian, $\eta$ and $\xi$, we can also define,
following \cite{Assaniousssi:2014ota}, a parameter $\beta$ which is more convenient for measuring the departure of our geometry from its classical limit. This is defined by the expression
\begin{equation}\label{defs}
\beta\equiv\frac{\langle \widehat{H_0^{-1}a^4}\rangle}{\langle \widehat{H_0^{-1}a^6}\rangle^{2/3}\langle \widehat{H_0^{-1}}\rangle^{1/3}}-1=\frac{\xi}{\eta^{2/3}}-1.
\end{equation}
It is trivial to see that $\beta=0$ when the above derived classicality condition $\xi=\eta^{2/3}$ is met.
Furthermore the expansion of $\bar{a}^2$ around $\vec{k}=\vec{0}$  (equal to the non-relativistic expansion given in \cite{Assaniousssi:2014ota}) is
\begin{equation}\label{scalefexpir}
\bar{a}^2(\vec{k}^2/m^2)\approx \eta^{1/3}\left[1+\frac{\beta}{3}\left(\frac{\vec{k}/\eta^{1/6}}{m}\right)^2\right].
\end{equation}
Thus we see in the low momentum limit $\vec{k}/m\ll1$, the parameter $\beta$ again measures the deviation from classicality.

\section{Dispersion relation on a quantum, cosmological spacetime}\label{disprelsection}

The dispersion relation for a $k$-mode of a scalar field with mass $m$ in a FRLW space-time is determined by the equations of motion of the Hamiltonian \eqref{quantumHamiltonianforkmodeFRW} in the classical limit. Since the relation \eqref{constraintpartialkNa3} holds good, the equations of motion are
\begin{align}
\dot{\phi}_{\vec{k}}=\,\frac{\bar{N}}{{a}^3}\,\pi_{\vec{k}},\quad \dot{\pi}_{\vec{k}}=\,-\phi_{\vec{k}}\,\frac{\bar{N}}{\bar{a}^3}\,(\vec{k}^2 \,\bar{a}^4 + \,m^2 \,\bar{a}^6).
\end{align}
Therefore, one has the constraint
\begin{equation}\label{eom}
\ddot{\phi}_{\vec{k}} =\,\bigg(-3\,H+\frac{\dot{\bar{N}}}{\bar{N}}\bigg)\,\dot{\phi}_{\vec{k}}-\bar{N}^2\,\phi_{\vec{k}}\,\bigg(\frac{\vec{k}^2}{\bar{a}^2} + \,m^2\bigg),
\end{equation}
where we use the definition of the Hubble rate $H={\dot{\bar{a}}}/{\bar{a}}$.\\
In the eikonal approximation ($\dot{N},\dot{a}\ll\omega_{\vec{k}}$) the dispersion relation of a mode $k$ reads as\footnote{The same relation can be obtained from the constraint $g_{\mu\nu}k^\mu k^\nu=-m^2$ valid for a point-like particle with time-like four-momentum.}
\begin{equation}
\frac{k_0^2}{\bar{N}^2}=\frac{\vec{k}^2}{\bar{a}^2}+m^2.
\end{equation}

We now introduce a comoving cosmological observer whose four-velocity is $u^\mu=(1/\bar{N}_{obs},0,0,0).$\footnote{We can be approximately described as comoving observers given that the peculiar velocity of earth is only about 360 Km/s with respect to the CMB frame.} This observer will experience a metric given by the following line element
\begin{equation}\label{obs}
\bar{g}_{\mu\nu}^{obs} dx^\mu dx^\nu=-\bar{N}_{obs}^2 dt^2+\bar{a}_{obs}^2\left(dx^2+dy^2+dz^2\right).
\end{equation}
Then the energy and the momentum of the particle measured by the observer are the following
\begin{equation}
E=u^\mu k_\mu=\frac{k_0}{\bar{N}_{obs}},\;\vec{p}^{\,2}=\left(\bar{g}^{\mu\nu}_{obs}+u^\mu u^\nu\right)k_\mu k_\nu=\frac{\vec{k}^2}{\bar{a}_{obs}^2},
\end{equation}
The on-shell relation written in terms of the $(E,\vec{p})$ variables reads as
\begin{equation}
-m^2=-f^2 E^2+g^2 \vec{p}^{\,2},
\end{equation}
where the rainbow functions are defined as $f:=\frac{\bar{N}_{obs}}{\bar{N}}$ and $g:=\frac{\bar{a}_{obs}}{\bar{a}}$ \cite{Assaniousssi:2014ota}. The previous relation can be rewritten as
\begin{equation}\label{disprel}
E^2=\frac{\bar{a}^4}{\bar{a}_{obs}^4}\vec{p}^{\,2}+\frac{\bar{a}^6}{\bar{a}_{obs}^6}m^2.
\end{equation}
By means of the equation in \eqref{equ_scfac} and the definition of $\delta$ in terms of $\eta$ and $\xi$, it follows the relation
\begin{equation}
m^2\,\frac{\bar{a}^6}{\bar{a}^6_{obs}}=\,m^2\Big(-\frac{\vec{k}^2}{m^2}\,\frac{\bar{a}^4}{\bar{a}^6_{obs}}
+\,\frac{\vec{k}^2}{m^2}\frac{\xi}{\bar{a}^6_{obs}}+\,\frac{\eta}{\bar{a}^6_{obs}}\Big).
\end{equation}
The relation \eqref{disprel} can then be written as
\begin{equation}
E^2=\frac{\left[\bar{a}_{obs}^2 \,c^2 \,\vec{p}^{\,2} (1+\beta)+\,c^4\,m^2 \eta^{1/3}\right] \eta^{2/3}}{\bar{a}_{obs}^6}.
\end{equation}
At this point we notice that the comoving observer has zero or negligible peculiar velocity and accordingly, making use of \eqref{scalefexpir}, one can use the relation $\bar{a}^6_{obs}=\eta$, which says that comoving frame has an associated scale-factor that is $\vec{k}$-independent.

The dispersion relation \eqref{disprel}  then becomes
\begin{equation}\label{disprel2}
E^2= \,m^2\,c^4 +(1+\beta)\,\vec{p}^{\,2}\,c^2,
\end{equation}
where we have explicitly introduced $c$ for the present (classical) value of the speed of light in vacuum.

Remarkably, in our quantum gravity derived geometry the dispersion relation \eqref{disprel2} is indeed exact, since no further assumptions than the ones defining the model as realized in \cite{Assaniousssi:2014ota} have been used in deriving this relation\footnote{In this work we take the point of view that since the peculiar velocity of the observer is negligible, we use \eqref{scalefexpir} to justify the choice of a $\vec{k}$-independent scale-factor for such an observer. One might wonder whether the choice of a non-comoving frame would bring additional corrections to \eqref{disprel2}. This is probably the case. However, these corrections would be proportional to the velocity of the frame and not to the particle momentum and as such would not be intrinsic to the particle physics.}. Thus, \eqref{disprel2} is valid not only in the low energy limit $|\vec{k}|/m\ll1$ as suggested in \cite{Assaniousssi:2014ota} but also in the high energy limit $m/|\vec{k}|\to 0$ and it does not show corrections proportional to higher power of the momentum.

\section{Phenomenology of the rainbow dispersion relation}\label{pheno}

Having now derived an exact form of the dispersion relation induced by the deviation from classicality of the effective space-time geometry, it is possible to address the kind of phenomenological constraints which can be derived in this context. In this sense we have to take into account two facts. First, the above derivation was done for a scalar field in a cosmological (homogeneous and isotropic) setting. We do not know how this treatment generalises to different background geometries and different fields. For the first point we shall hence avoid statements based on ground based experiments or particle phenomenology in strong gravity environments where the isotropy and homogeneity requirements for the metric will fail. Second, the treatment as proposed in [1] clearly differentiates between quantum gravitational and quantum matter degrees of freedom (the former being traced out). If the quantum nature of the fields is a necessary requirement for seeing quantum features of space-time then when dealing with low-energy classical gravitational waves one can assume that they would propagate at the standard speed of light $c$.

Given these facts one can envisage three possible scenarios:
\begin{itemize}
\item
{\bf Scenario 1: field dependent $\beta$}.
Let us assume for the moment that the parameter of deviation from classicality $\beta$ depends on the type of field considered.
Even sticking to cosmological constraints this would severely constraint the parameter. Just to make an example if light and electrons have different limit speeds the constraints on the relative difference  would be  $(|\Delta c^2|/c^2)_{\gamma,e}= O(10^{-16})$ just using the absence of gamma decay in electron-positron pair for the 80 TeV photons reaching us from the Crab nebula and the deduced at least comparable energy for the electrons responsible of such photons via inverse Compton scattering \cite{Liberati:2013xla} (the constraint can be reduced to $O(10^{-15})$ if one prefers to use 20 TeV gamma rays reaching us from cosmological distances, e.g. from an active galactic nuclei as Markarian 501~\cite{Coleman:1998ti}).

\item
{\bf Scenario 2: Universal $\beta$}.
The expression \eqref{defs} shows that the parameter $\beta$  only depends on the quantum gravity state $\psi\in\,\mathcal{H}_G$. This provides an argument in favour of the universality of the expression \eqref{disprel2}. However, universality of $\beta$ has strong consequences for the phenomenology of \eqref{disprel2}.
In this case the only observable effects would be associated to the interaction of matter with gravity\footnote{Within the matter sector a universal $\beta$ could be reabsorbed by a simple redefinition of the speed of light and of the particle masses.} which, as said before, could be unaffected by the gravitational quantum correction and hence characterized by the geometric limit speed $c$ as set by the classical limit of the metric.
Therefore if $\beta>0$, one would naturally expect that particles will be able to rapidly loose energy via vacuum \v{C}erenkov radiation of gravitons. Constraints from the observation of ultra high energy cosmic rays where derived in \cite{Moore:2001bv}. The rate of energy loss was calculated to be
\begin{equation}
\frac{dE}{dt}=\frac{G p^{4}}{3}(c_p-c)^2
\end{equation}
where $c_p=c\sqrt{1+\beta}$ is the speed of the particle and $G$ is Newton's constant. The corresponding constraint from the observation of high energy cosmic rays is $\beta \lesssim O(10^{-15})$. This bound assumes that the cosmic rays are protons, uses the highest record energy $3 \cdot 10^{20}$ eV, and assumes that the protons have traveled over at least 10 kpc.
Assuming (as normally given) that these cosmic ray reach us from extra-galactic distances improves the limit up to $\beta \lesssim O(10^{-19})$~\cite{Moore:2001bv}.

If $\beta$ is taken to be negative, although bounded from below ($\beta\geq -1$) assuming $\eta$ and $\xi$ positive, then the speed of gravity is higher than the limit speed of massive particles. In this case one can still put a bound on the value of the parameter considering emission of photons from gravitational waves through ordinary \v{C}erenkov effect, but it is much less stringent being based on precision tests of General Relativity. In this context the prediction of the orbital decay of binary pulsars agrees with $c_g=c$ at $1\%$ \cite{Carlip:1999an,Taylor:1993ni}.

\item
{\bf Scenario 3: Time dependent $\beta$}. Within a cosmological context it is still conceivable that the classicalization of the universe has progressed in time starting from a relatively large value of $\beta$ which has been driven towards zero with cosmological expansion. This hypothesis would require some plausible argument for the initial value of $\beta$ and his evolution in time. Determining such evolution is beyond the scope of this work but one can easily foresee that a varying $\beta$ could be easily be responsible for interesting phenomenology. For example if a universal (except for gravity) $\beta(t)$ had a dramatic transition from approximately one (quantum phase) to zero (classical phase) in the early universe this would reproduce the basic setting for bi-metric varying speed of light scenario (see e.g.~\cite{Bassett:2000wj}) which can lead to a spectrum of primordial perturbations and a resolution of the horizon problem.
From the point of view of phenomenological constraints this scenario could be tested via future observation of primordial gravitational waves imprint in the B modes of the cosmic microwave background (as the scalar to tensor ratio would be modified if the speed of light/inflaton and gravity do not coincide). In the most general case that different matter field are endowed with different $\beta$ with non-negligible differences in the early universe, constraints on $\Delta\beta$ could be provided by Big Bang nucleosynthesis and CMB observations (e.g.~via possible modifications of the physic at recombination).
\end{itemize}

\section{Conclusions}\label{concl}

In conclusion one can say that the framework developed in \cite{Ashtekar:2009mb,Assaniousssi:2014ota} and in the present paper offers a very general prediction for a regime, stemming from a quantum treatment of the gravitational sector, where a continuous space-time has emerged but might retain a quantum nature which can be probed by quantum fields. In this paper we have shown that an exact treatment of the emergent rainbow ($\vec{k}$-depedent) geometry lead, surprisingly, to a dispersion relation which is of relativistic form albeit characterized by a shift in the limit speed of propagation. So if (as it seems to be hinted by the present derivation) the non-classicality parameter $\beta$ is universal for any non-gravitational field, then no Lorentz violation will be detectable in the matter sector of the Standard Model and only gravity-matter phenomenology might be able to show any deviation from standard physics.

Of course, the simplicity of the present model does not allow to have sharp predictions for what regards the phenomenology associated to this approach. Still we have seen that present constraints on the value of $\beta$ are already very stringent both in the first two of the above discussed scenarios (if $\beta\geq 0$). This seems to hint that if anything like this framework is realised in nature then a ``classicalization" of space-time should be very rapidly achieved following the continuous limit. It would be hence important to develop this model in the sense of being able to predict some time  evolution of the $\beta$ factor in the early universe. This in turn could allow to discuss interesting phenomenology (e.g. varying speed of light scenarios) and cast constraints.
Finally, this framework is presently developed only for cosmological solutions. It would be interesting to extend it to more general background as for example very high precision ground based experiments with quantum objects (such as Bose-Einstein condensates, see e.g.~\cite{AmelinoCamelia:2009zzb}) could be more appropriate to test this kind of tiny deviations from classical space-time.

\acknowledgments 
The authors wish to thank Andrea Dapor for his comments on a draft version of the paper.
R. G. T. was financially supported by PNPD-CAPES n. 2265/2011, Brazil. 
S.L. and M.L. acknowledge financial support from the John Templeton Foundation (JTF), grant \#51876.

\appendix
\section{Massless fields}\label{app}
Let us apply the procedure reviewed in Sec.\ref{cosmst} to a massless field $\theta$. The constraints \eqref{equ_scfac} read as
\begin{equation}
\frac{\bar{N}_\theta}{\bar{a}_\theta^3}=\,\langle \hat{H}^{-1}\rangle,\quad \frac{\bar{N}_\theta}{\bar{a}_\theta^3}\,\vec{k}^2\,\bar{a}_\theta^4=\,\langle\hat{\Omega}(\vec{k},m=0)\rangle.
\end{equation}
In this case it is possible to solve for  $\bar{a}^2_\theta$ straightforwardly from the second expression,
\begin{equation}
\bar{a}_\theta^2=\,\frac{1}{|\vec{k}|}\,\Big(\frac{\langle \hat{\Omega}(\vec{k},m)\rangle}{\langle \hat{H}^{-1}_0\rangle}\Big)^{1/2}.
\end{equation}
Since for $m=0$ one has (see \eqref{omega})
\begin{equation}
\hat{\Omega}(\vec{k},m=0)=\,\vec{k}^2\,\widehat{H^{-1}_0a^4},
\end{equation}
it follows that
\begin{equation}
\langle\hat{\Omega}(\vec{k},m=0)\rangle = \,\vec{k}^2\,\langle \widehat{H^{-1}_0a^4}\,\rangle=\vec{k}^2\,\langle \hat{H}^{-1}_0\rangle\,\xi.
\end{equation}
Hence the scale-factor $\bar{a}^2_\theta$ and the lapse time function $\bar{N}_\theta$ are independent of $\vec{k}$,
\begin{equation}\label{aNform=0}
\bar{a}_\theta^2=\,\Big(\frac{\langle \widehat{H^{-1}_0a^4}\rangle}{\langle \hat{H}^{-1}_0\rangle}\Big)^{1/2}=\sqrt{\xi},\,\quad \bar{N}_\theta=\,\big(\langle \widehat{H^{-1}_0a^4}\rangle\big)^{3/4}\,\langle \hat{H}^{-1}_0\rangle^{1/4}=\xi^{3/4}\,\langle \hat{H}^{-1}_0\rangle.
\end{equation}
Therefore particles with different $\vec{k}$ experience the same scale-factor and hence the same metric. Moreover since from \eqref{aNform=0} it can be seen that they carry information only about the parameter $\xi$ and not $\eta$, it is not possible to reconstruct $\beta$ using only massless particles. Hence even if space-time has quantum feature ($\beta\neq 0$), massless particles would see it as classical.

Note that in the classical limit $\beta\rightarrow 0$, one can write
\begin{align}\label{classicalimit1}
&\xi=\frac{\langle \widehat{H_0^{-1}a^4}\rangle}{\langle \widehat{H}_0^{-1}\rangle}\rightarrow a_{cl}^4,\\\label{classicalimit2}
&\eta=\frac{\langle \widehat{H_0^{-1}a^6}\rangle}{\langle \widehat{H}_0^{-1}\rangle}\rightarrow a_{cl}^6,
\end{align}
and the metric defined by \eqref{aNform=0}  becomes the ordinary FRW flat metric given by the following line element
\begin{equation}\label{classicalmetric}
ds^2=g^{(cl)}_{\mu\nu}dx^\mu dx^\nu=-N^2_{cl} dt^2+a^2_{cl}(dx^2+dy^2+dz^2).
\end{equation}
By the equivalence principle (and the absence of back-reaction) \eqref{classicalmetric} has to be the same metric given by the classical limit of the metric experienced by a massive field. This is indeed the case given that in the limit $\beta\rightarrow 0$ the general rainbow scale factor $\bar{a}^2$ introduce in \eqref{solutions} for a massive scalar field reduces to  $\bar{a}^2(\vec{k},m,\xi,\eta)\rightarrow \bar{a}^2_0=\sqrt{\xi}=\eta^{1/3}$ giving then the classical scale-factor by means of \eqref{classicalimit1}\eqref{classicalimit2}.

\end{document}